\documentclass[12pt]{article}
\usepackage{graphicx}
\usepackage{epsfig}
\usepackage[figuresright]{rotating}
\usepackage{epstopdf}
\usepackage{amsmath}
\usepackage{amssymb}
\usepackage{amsfonts}

\usepackage{cite}
\begin{document}
\renewcommand{\thefootnote}{\fnsymbol{footnote}}
\begin{center}
{\Huge \bf Polarizability of the nucleon}\\
[1ex] 
Martin Schumacher\footnote{mschuma3@gwdg.de}\\
II.  Physikalisches Institut der Universit\"at G\"ottingen,
Friedrich-Hund-Platz 1\\ D-37077 G\"ottingen, Germany
\end{center}

\begin{abstract}
The status of the experimental and theoretical  investigations on the 
polarizabilities of the nucleon is presented. This includes a  confirmation 
of the validitiy of the previously introduced {\it recommended}  values 
 of the polarizabilities \cite{schumacher05,schumacher13}.
It is shown that the most reliable  approach
to a prediction of the polarizabilities is obtained from the nonsubtracted 
dispersion theory, where the appropriate degrees of freedom taken from other
precise experimental data are taken into account. The present values of the
{\it recommended}
polarizabilities are $\alpha_p=12.0\pm 0.5$, $\beta_p=1.9\mp 0.5$,
$\alpha_n=12.6\pm 1.2$, $\beta_n= 2.6 \mp 1.2$ in units of $10^{-4}$fm$^3$
and $\gamma^{(p)}_\pi=-36.4\pm 1.5$, $\gamma^{(n)}_\pi= +58.6\pm 4.0$,
$\gamma^{(p)}_0=-0.58\pm 0.20$, $\gamma^{(n)}_0=+0.38\pm 0.22$ in units of
$10^{-4}$fm$^4$.
\end{abstract}

\renewcommand{\thefootnote}{\arabic{footnote}}
\section{Introduction}

The polarizabilities belong to the fundamental structure constants of the
 nucleon, in addition to the mass, the electric charge, the spin and the
 magnetic moment. The proposal to measure the polarizabilities dates back to
 the 1950th. Two experimental options were considered (i) Compton scattering
 by the proton and (ii) the scattering of slow neutrons in the Coulomb field
 of heavy nuclei. The idea was that the nucleon with its ``pion cloud'', i.e.
pions being part of the constituent-quark structure, obtains
 an electric dipole moment under the action of an electric field vector which
 is proportional to the electric polarizability. After the discovery of the
 photoexcitation of the $\Delta$ resonance it became obvious 
that the nucleon also
 should have a strong paramagnetic polarizability, because of a virtual
 spin-flip transition of one of the constituent quarks due to the magnetic
 field vector provided by a real photon in a Compton scattering experiment.
 However, experiments showed that this expected strong paramagnetism is not
 observed. Apparently a strong diamagnetism exists which compensates the
 expected strong paramagnetism. Though this explanation is straightforward,
 it remained unknown how it may be understood in terms of the structure of the
 nucleon \cite{schumacher05}. A solution of this problem was found  later
when it was shown
 that the diamagnetism is a property of the structure of the constituent
 quarks \cite{levchuk05,schumacher06,schumacher07a,schumacher07b,schumacher08}. 
In retrospect this is not a surprise, 
because constituent quarks
 generate
 their mass mainly through interaction with the QCD vacuum via the exchange
 of a $\sigma$  meson. This mechanism is predicted by the linear $\sigma$
 model on the quark 
 level (QLL$\sigma$M) \cite{schumacher13} which also predicts 
the mass of the $\sigma$ meson to be
 $m_\sigma$=666 MeV. 
 The $\sigma$  meson has the capability of interacting with two photons being in
 parallel planes of linear polarization. We will show in the following that
 the $\sigma$ meson as part of the constituent quark structure, therefore, 
provides 
the largest part of the electric polarizability and the total diamagnetic
 polarizability.

\section{Definition of electromagnetic polarizabilities}

A nucleon in an electric field {\bf E} and a magnetic field {\bf H} obtains 
an electric dipole moment {\bf d} and magnetic dipole moment {\bf m} given 
by \cite{schumacher05}
\begin{eqnarray}
{\mathbf d}\,\,=4\pi\,\alpha \,{\mathbf E}\\
{\mathbf m}=4\pi\,\beta \,{\mathbf H}
\end{eqnarray}
in a unit system where the electric charge $e$ is given by
$e^2/4\pi=\alpha_{em}=1/137.04$. The proportionality constants 
${\alpha}$ and ${ \beta}$ are denoted as the electric and
 magnetic polarizabilities, respectively. These polarizabilities may be
 understood
as a measure of the response of the nucleon structure to the fields provided
 by a real or virtual photon and it is evident that we need a second photon
 to measure  the polarizabilities. This may be expressed through the relations
\begin{equation}
\delta W=-\frac12 \,4\pi\,\alpha\,{\mathbf E}^2-\frac12
\,4\pi\,\beta\,{\mathbf H}^2
\end{equation}
where $ \delta W$ is the energy change in the electromagnetic field due 
to the presence of the nucleon in the field. The definition implies that 
the polarizabilities are measured in units of a volume, i.e. in units of 
fm$^3$ (1 fm=$10^{-15}$ m).

\section{Modes of two-photon reactions and experimental methods}

Static electric fields of sufficient strength are provided by the Coulomb
 field of heavy nuclei. Therefore, the electric polarizability of the neutron
can be measured by scattering slow neutrons in the electric field {\bf E} of
 a Pb nucleus. The neutron has no electric charge.  Therefore,
two simultaneously interacting electric field vectors (two virtual photons)
 are required to produce a deflection of the neutron. Then the electric 
polarizability can be obtained from the differential cross section measured 
at a small
deflection angle. A further possibility is provided by Compton scattering of 
real photons by the nucleon, where during the scattering process two electric 
and two magnetic field vectors simultaneously interact with the nucleon.  

In the following we discuss the experimental options we have to measure the 
polarizabilities of the nucleon. As outlined above two photons are needed 
which simultaneously interact with the electrically charged parts of the 
nucleon.
These photons may be in parallel or perpendicular planes of linear 
polarization and in these two modes measure the
polarizabilities $\alpha$, $\beta$ or 
spinpolarizabilities $\gamma$, respectively. 
The spinpolarizability is nonzero only for particles having a spin. 

In total the experimental options discussed above provide us with 6
combinations of two electric and magnetic field vectors. These are described
in the 
following two equations:

$\bullet$ For photons in parallel planes of linear polarization we have
\begin{equation}
(\text{case}\, 1)\,\, \alpha: \,\,\,\,{\mathbf E}\uparrow\uparrow 
{\mathbf E}'\quad\quad (\text{case}\, 2)\,\,\beta: \,
{\mathbf H} \rightarrow\rightarrow {\mathbf H}'\quad\, (\text{case}\,\, 3)
\,\, -\beta: \,{\mathbf H}\rightarrow\leftarrow {\mathbf H}'\label{def1}
\end{equation}

$\bullet$ For photons in perpendicular planes of linear polarization we have
\begin{equation}
(\text{case}\,\,4)\,\, \gamma_E: {\mathbf E}\uparrow\rightarrow 
{\mathbf E}'\quad \,\,(\text{case}\,\,5)\,\,\gamma_H: {\mathbf H} 
\rightarrow\downarrow {\mathbf H}'\quad \,\,(\text{case}\,\,6)\,\, -\gamma_H:
 {\mathbf H}\rightarrow\uparrow {\mathbf H}'\label{def2}
\end{equation}

\noindent
Case (1) corresponds to the measurement of the electric polarizability 
$\alpha$ via two parallel electric field vectors {\bf E} and {\bf E'}. 
These parallel electric field vectors may either be provided as longitudinal
photons by the Coulomb field of a heavy nucleus, or by Compton scattering 
in the forward
direction or by reflecting the photon by $180^\circ$. Real 
photons simultaneously
provide transvers electric {\bf E} and magnetic {\bf H} field vectors. This
means that in a Compton scattering experiment linear combinations of electric
and magnetic polarizabilities and linear combinations of electric and 
magnetic 
spinpolarizabilities are measured.
The combination of case (1) and case (2) measures $\alpha+\beta$ 
and is observed in forward-direction Compton scattering. The combination
of case (1) and case (3) measures $\alpha-\beta$ and is observed
 in backward-direction Compton scattering.The combination of case (4) and
 case (5)
measures $\gamma_0\equiv \gamma_E + \gamma_H$ and is observed in
 forward-direction Compton scattering.
The combination of case (4) and case (6)
measures $\gamma_\pi\equiv \gamma_E-\gamma_H$ and is observed in
 backward-direction Compton scattering.
Compton scattering experiments exactly in the forward direction and exactly
 in the backward direction are not possible
from a technical point of view. Therefore, the respective quantities have 
to be extracted from Compton scattering experiments carried out at 
intermediate angles.

\section{Experimental results}

The experimental polarizabilities of the proton (p) and the neutron (n) 
may be summarized as follows
\begin{equation}
\alpha_p=12.0\pm 0.5,\quad \beta_p=1.9\mp 0.5, \quad 
\alpha_n=12.6\pm 1.2, \quad \beta_n=2.6\mp 1.2
\end{equation}
{  in\, units\, of \,\,$ 10^{-4} \,{\rm fm}^3 $.\\

\noindent
The experimental spinpolarizabilities of the proton (p) and neutron (n) are
\begin{equation}
\gamma^{(p)}_\pi=-36.4\pm 1.5, \quad \gamma^{(n)}_\pi=58.6\pm 4.0 \quad
\text{ in units of}\,\, 10^{-4}\,{\rm fm}^4.
\end{equation}

The experimental polarizabilities of the proton have been obtained as an 
average from a larger number of Compton scattering experiments
\cite{schumacher05}. In addition a recent reanalysis of these data 
leading to $\alpha_p=12.03\pm 0.72$
has been
taken into account \cite{pasquini19}. 
The 
experimental electric polarizability of the neutron is the average of an 
experiment on electromagnetic scattering of a  neutron in the Coulomb field 
of a Pb nucleus and a Compton scattering experiment on a quasifree  neutron, 
i.e. a neutron
separated from a deuteron during the scattering process. The two results are  
\cite{schumacher05} 
$\alpha_n=12.6\pm 2.5$ from electromagnetic  scattering of a slow 
neutron in the electric field of a Pb nucleus, and
$\alpha_n=12.5\pm 2.3$ from quasifree Compton scattering by a 
neutron initially bound in the deuteron. In addition the result obtained from
the experimental electric polarizability of the proton $\alpha_p$ 
and the predicted ratio
$\alpha_n/\alpha_p$ leading to $\alpha_n=12.7\pm 0.9$ 
has been taken in account \cite{schumacher13a}.
The average given above is obtained from these three  numbers.

Furthermore, there have been  experiments at the University of Lund 
(Sweden) where the electric polarizability of the neutron is determined 
through Compton scattering by the deuteron. The results obtained in this way
are model dependent.

\section{Calculation of polarizabilities}

Recently great progress has been made in disentangling the total 
photoabsorption cross section into parts
separated by the spin, the isospin and the parity of the intermediate 
state \cite{schumacher09,schumacher11}}, using the meson photoproduction 
amplitudes of 
Drechsel et
al.\cite{drechsel07}
 The spin of the intermediate state 
may be $s=1/2$ or $s=3/2$ depending on the spin 
directions of the photon and the nucleon in the initial state.
The parity change during the transion from the ground state to the 
intermediate state is $\Delta P= \text{yes}$ for the multipoles 
$E1,\,M2,\cdots$ and $\Delta P= \text{no}$ for the 
multipoles 
$M1,\,E2,\cdots$. Calculating the respective partial cross 
sections from photo-meson data, the following
sum rules can be evaluated:
\begin{eqnarray}
&&\!\!\!\!\!\!\alpha+\beta=\frac{1}{2\pi^2} \int^\infty_{\omega_0}
\frac{\sigma_{\rm tot}(\omega)}{\omega^2}d\omega, \label{e1}\\
&&\!\!\!\!\!\! \alpha-\beta=\frac{1}{2\pi^2}\int^\infty_{\omega_0}
\sqrt{1+\frac{2\omega}{m}}\left[\sigma(\omega,E1,M2,\cdots)-
\sigma(\omega,M1,E2,\cdots)\right]\frac{d\omega}{\omega^2} +
(\alpha-\beta)^t \label{e2}\\
&&\!\!\!\!\!\!\gamma_0=-\frac{1}{4\pi^2}\int^\infty_{\omega_0}
\frac{\sigma_{3/2}(\omega)-\sigma_{1/2}(\omega)}{\omega^3}d\omega,\label{e3}\\
&& \!\!\!\!\!\!\gamma_\pi=\frac{1}{4\pi^2}\int^\infty_{\omega_0}
\sqrt{1+\frac{2\omega}{m}}\left(1+\frac{\omega}{m}\right)
\sum_n P_n[\sigma^n_{3/2}(\omega)-\sigma^n_{1/2}(\omega)]
\frac{d\omega}{\omega^3}+\gamma^t_\pi, \label{e4}\\
&&\!\!\!\!\!\!P_n=-1\,\text{for}\,E1,M2,\cdots\,\text{multipoles and}\, 
P_n=+1\,text{for} \,M1,E2,\cdots\,\text{multipoles}.\label{e5}\\
&&\!\!\!\!\!\!(\alpha-\beta)^t=\frac{1}{2 \pi}\left[\frac{g_{\sigma NN}{ M}
(\sigma\to \gamma\gamma)}{ m^2_\sigma}
+\frac{g_{f_0 NN}{ M}(f_0\to \gamma\gamma)}{ m^2_{f_0}}
+\frac{g_{a_0 NN}{ M}(a_0\to \gamma\gamma)}{
  m^2_{a_0}}\tau_3\right],\label{e6} \\
&&\!\!\!\!\!\!\gamma^t_\pi=\frac{1}{2\pi m}\left[
\frac{g_{\pi NN}{ M}(\pi^0\to \gamma\gamma)}{ m^2_{\pi^0}}\tau_3
+\frac{g_{\eta NN}{ M}(f_0\to \gamma\gamma)}{m^2_\eta}
+\frac{g_{\eta' NN}{ M}(\eta'\to \gamma\gamma)}{m^2_{\eta'}}\right]\label{e7}.
\end{eqnarray}
where $\omega$ is the photon energy in the lab frame.
The sum rules for $\alpha+\beta$ and $\gamma_0$ 
depend on nucleon-structure degrees of freedom only, whereas the sum rules 
for $\alpha-\beta$ and $\gamma_\pi$ have to be 
supplemented by the quantities $(\alpha-\beta)^t$ and 
$\gamma^t_\pi$, respectively. These are $t$-channel
contributions which may be interpreted as contributions of  scalar and 
pseudoscalar mesons being parts of the constituent-quark structure. The 
sum rule for
$\alpha+\beta$  depends on the total photoabsorption cross 
section  and, therefore, does not require a disentangling with respect to 
quantum numbers.
The sum rule for $\alpha-\beta$  requires a disentangling with 
respect to the parity change of the transition. The sum rule for 
$\gamma_0$
requires a disentangling with respect to the spin of the intermediate state. 
The sum rule for $\gamma_\pi$ requires a disentangling with 
respect to
spin and parity change.

The $t$-channel contributions depend on those scalar and 
pseudoscalar mesons which (i) are part of the structure of the constituent 
quarks and (ii)
are capable of coupling to two photons. These are the mesons 
$\sigma(600)$, $f_0(980)$ and $a_0(980)$
in case of $(\alpha-\beta)^t$, and the mesons $\pi^0$, 
$\eta$ and $\eta'$ in case of 
$\gamma^t_\pi$.
The contributions are dominated by the $\sigma$ and the 
$\pi^0$ mesons  whereas the other mesons only lead to small corrections.

\section{Results of calculation}

The results of the calculation are summarized in the following ten
equations \cite{schumacher09,schumacher11}:
\begin{eqnarray}
&&\alpha_p=\,\,\,+4.5\,\text{(nucleon)}+7.6\,
\text{(const. quark)}=+12.1\label{e8}\\
&&\beta_p=\,\,\,+9.4\,\text{(nucleon)}-7.6\,
\text{(const. quark)}=\,\,\,+1.8\label{e9}\\
&&\alpha_n=\,\,\,+5.1\,\text{(nucleon)}+7.6\,\text{(const. quark)}=+12.7
\label{e10}\\
&&\beta_n=+10.1\,\text{(nucleon)}-7.6\,\text{(const. quark)}=\,\,\,
+2.5\label{e11}\\&&\text{in units of}\,\,10^{-4}{\rm fm}^3\\
&&\gamma^{(p)}_0=\,\,\,-0.58\pm 0.20\,\text{(nucleon)}\label{e12}\\
&&\gamma^{(n)}_0=\,\,\,+0.38\pm 0.22\,\text{(nucleon)}\label{e13}\\
&&\gamma^{(p)}_\pi=\,\,\,+8.5\,\text{(nucleon)}-45.1\,\,
\text{(const. quark)}=-36.6 \label{e14}\\
&&\gamma^{(n)}_\pi=+10.0\,\text{(nucleon)}+48.3\,\,\text{(const. quark)}
=+58.3\label{e15}\\
&&\text{in units of}\,\,10^{-4}{\rm fm}^4\label{e16}
\end{eqnarray}

 The electric polarizabilities $\alpha_p$ and $\alpha_n$ 
are dominated by a smaller component due to the pion cloud (nucleon)
and a larger component due to the $\sigma$ meson as part of the 
constituent-quark structure (const. quark). The magnetic polarizabilities
$\beta_p$ and $\beta_n$ have a large paramagnetic part 
due to the spin structure of the nucleon (nucleon) and an only slightly smaller
diamagnetic part due to the $\sigma$ meson as part of the 
constituent-quark structure (const. quark). The contributions of the 
$\sigma$
meson may be supplemented by small corrections due to $f_0(980)$ 
and $a_0(980)$ mesons 
\cite{schumacher07b,schumacher08,schumacher09,schumacher11}.
 These contributions are disregarded here because of their smallness and
uncertainties \cite{schumacher13a}.

The spinpolarizabilities $\gamma_0^{(p)}$ and 
$\gamma_0^{(n)}$ are dominated by destructively interfering 
components from the pion cloud
and the spin structure of the nucleon. The different signs obtained for 
the proton and the neutron are due to this destructive interference.
\cite{schumacher11}
The spinpolarizabilities $\gamma^{(p)}_\pi$ and 
$\gamma^{(n)}_\pi$ have a minor component due to the structure 
of the nucleon (nucleon)
and a major component due to the pseudoscalar mesons $\pi^0$, 
$\eta$ and $\eta'$ as structure components of the 
constituent quarks
(const. quark).  

Differing from other theoretical approaches the presently applied 
dispersion theory is based on fundamental relations only. The precision
of the results of the present calculation only depends  on the precision of
the photomeson data used as an input.  A consideration shows that the
errors of the  results given in Eqs. (\ref{e8}) - (\ref{e16})
are of the same order of magnitude as, or somewhat smaller than
those of the corresponding experimental
results. 

\section{Discussion}

In a first approach the electric polarizabilities of proton and neutron
have been related to the dipole moment of the transitions 
\begin{center}
\begin{equation}
p\to n + \pi^+ \quad {\rm and} \quad  n \to p + \pi^-.
\end{equation}
\end{center}
Since the $n + \pi^+$ dipole moment is smaller than the $p + \pi^-$ dipole
moment we expect that the related contributions to the electric polarizabilities
in Eq. (\ref{e8}) and (\ref{e10}) are smaller  for the proton than for the
neutron. This is in agreement with the observation, where $\alpha_p$(nucleon)
=+4.5 (Eq. \ref{e8}) and $\alpha_n$(nucleon)=+5.1 (Eq. \ref{e10}) are given. The
difference between the two numbers 
{\it viz.} $\alpha_n$(nucleon)-$\alpha_p$(nucleon) = 0.6 precisely
corresponds to the difference beteen the electric polarizabilities of neutron
and proton, as seen in Eq. (\ref{e8}) and (\ref{e10}). The reason for this
agreement is that the constituent-quark parts of the two polarizabilities
are the same.
 
The quantity $7.6$(const. quark) entering into Eqs. (\ref{e8}) to
(\ref{e11}) corresponds to the $\sigma$ meson as part of the constituent-quark
structure. This quantity has a positive sign when being part of the electric
polarizabilities, or a negative sign when representing the diamagnetic
polarizabilities. The investigation of these quantities has been carried out 
previously in a number of publications
\cite{schumacher10,schumacher11a,schumacher14,schumacher16,schumacher18}. 
All the relevant
information may be found in these publications.

The meaning of the spinpolizabilities in relation to the structure of the
nucleon is less straightforward than that of the polarizabilities.

In addition to dispersion theory chiral perturbation theory plays a prominent
r\^ole in current investigations of nucleon Compton scattering and
polarizabilities. Therefore, it is advisable  to carry out a comparison of the
two approaches. This is done in the following table:

\begin{table}[h]
\begin{center}
\caption{Predicted electric $\alpha_p$ and magnetic $\beta_p$ polarizabilities
for the proton,
where BChPT \cite{lensky15} denotes covariant chiral perturbation theory and
DR dispersion  
theory}
\vspace{10mm}
\begin{tabular}{c||c|c||c|c||c|c||c|c||}
\hline\hline
&$\alpha_p$(BChPT)&$\alpha_p$(DR)&$\beta_p$(BChPT)&$\beta_p$(DR)\\
\hline
N$\pi$&+6.9&+3.09&-1.8&+0.48\\
$\Delta \pi$&+4.4&+1.4&-1.4&+0.4\\
$\Delta$-pole&-0.1&-0.01&+7.1&+8.56\\
t-channel&--&+7.6&--&-7.6\\
Total&+11.2&+12.1&+3.9&+1.8\\
\hline\hline
\end{tabular}
\end{center}
\end{table}

One essential difference between the two versions is  the missing $t$-channel
contribution in the BChPT version. The $t$-channel provides the total
diamagnetism and the largest part of the electric polarizability. An other
essential difference is contained in the N$\pi$ component of the
electric polarizability. The procedure used in the BChPT method 
corresponds to the Born approximation of the DR method, leading to an
error in the BChPT method of more than a factor of 2.

\end{document}